\documentclass[aps,superscriptaddress,eqsecnum,nofootinbib,showpacs,preprintnumbers]{revtex4}
\usepackage{amsmath}
\usepackage{amssymb}
\usepackage{amsfonts,amsthm,bm}
\usepackage{color,xcolor}
\usepackage[greek,english]{babel}
\usepackage{graphicx,epsfig}
\graphicspath{{figures/}{./}}
\usepackage{float}
\usepackage{subfigure}
\usepackage{tikz}
\usepackage[pdfencoding=auto,colorlinks=true,
allcolors=blue]{hyperref}

\newcommand{\n}{\nu}
\newcommand{\Om}{\Omega}

\newcommand{\g}{\gamma}
\newcommand{\abs}[1]{\left| #1 \right|}

\newcommand{\be}{\begin{eqnarray}}
\newcommand{\ee}{\end{eqnarray}}
\newcommand{\bea}{\begin{eqnarray}}
\newcommand{\nn}{\nonumber}
\newcommand{\eea}{\end{eqnarray}}

\def\g{\gamma}

\def\n{\nu}

\begin{document}

\title{Wormhole Solutions in Modified Brans-Dicke Theory}


\author{Eleftherios Papantonopoulos}
\email{lpapa@central.ntua.gr} \affiliation{Physics Division,
National Technical University of Athens, 15780 Zografou Campus,
Athens, Greece}

\author{Christoforos Vlachos}
\email{ cvlach@mail.ntua.gr}
\affiliation{Physics Division, National Technical University of
Athens, 15780 Zografou Campus, Athens, Greece}

\begin{abstract}

We  consider  a modified  Brans-Dicke theory  in which except the usual Brans-Dicke parameter a new dimensionful  parameter  appears which modifies the kinetic term of the scalar field  coupled to   gravity. Solving the coupled Einstein-Klein-Gordon equations
we find new  spherically symmetric  solutions. Depending on the choices of the parameters these solutions reduce to the  Schwarzschild solution of General Relativity and they give new wormhole solutions which depend on the new parameter.

\end{abstract}

\maketitle

\section{Introduction}
\label{Sec1}

Scalar fields play an important role in the General Relativity (GR) on short and large distances. On short distances
they are dressing the local black hole solutions with scalar hair and they provide wormhole solutions while on large distances
they describe the early inflationary  universe and also its late-times cosmic evolution. In a attempt to provide
a viable theory of gravity and to cure certain inconsistencies of GR, scalar-tensor theories were introduced.
As it is well known, Brans-Dicke  theory (BD) \cite{Brans:1961sx} is one of the first scalar-tensor gravity theories that modifies  GR
 in a viable way and respects Mach's principle and weak equivalence principle (WEP). In this theory there is an effective  Newtonian gravitational
constant $G$ which is the  inverse of the scalar field, $G\sim\frac{1}{\phi}$. It is characterized by a new dimensionless coupling constant $\omega$
large values of which  mean a significant contribution from the tensor part, while  scalar field contribution is important for small values.
 GR is recovered in
the limit $\omega\rightarrow \infty$.

It is interesting to note that BD theory     appears in
supergravity models such as in string theory at low-energies  or in the   Kaluza-Klein theories
after a dimensional reduction process \cite{Freund:1982pg}. These  theories  yield the correct Newtonian weak-field limit,
but care should be taken when one studies these theories and compare their predictions with GR. In general scalar fields, depending on their coupling to gravity, mediate fifth forces. In the case of BD solar system measurements of
post-Newtonian corrections require that $\omega$ is larger than a few thousands \cite{Bertotti:2003rm}.
Therefore in these theories scalar fields should accommodate a mechanism to suppress the scalar interaction on small scales.
 There are various  screening mechanisms to suppress scalar interactions on small scales. One of the basic screening mechanism is the Vainshtein mechanism \cite{Vainshtein:1972sx} which was developed for the massive gravity (for an extensive review on the Vainshtein mechanism in massive gravity see \cite{Babichev:2013usa}).

 On large scales, $\omega$ gets substantially lower values in a model dependent way \cite{Acquaviva:2004ti}, from cosmological observations.
 On the other side, the
gravitational coupling may depends on the scale \cite{Mannheim:1999bu}, having different value at local and at cosmological scale. In this case
$\omega$ can  be smaller at cosmological scales giving deviations from GR,
while agreement with local tests is preserved. Special solutions on BD cosmology have been given in \cite{nariai} which are generalizations
of the dust solution first given by Brans-Dicke. In \cite{Morganstern:1971dz} special radiation
solutions for spatially curved space were found and in \cite{O'Hanlon:1972hq} vacuum
solutions were given. In \cite{Gurevich} the general spatially flat cosmological
solution was obtained in parametric form for any barotropic perfect fluid.
In \cite{Petzold} the general stiff and radiation solutions for all kinds of spatial curvature
are found. Similar solutions were found in \cite{ruban}.
In \cite{Uehara:1981nq} exact solutions were found in the presence of cosmological constant.
Other works with exact solutions are given in \cite{Bertobarrow}. In \cite{Park:1997dw} it was argued  the initial singularity in the BD theory can be resolved.
BD theory is used to solve  some problems of the
inflationary scenario \cite{Mathiazhagan:1984vi}. Also a solution to the   ``graceful exit'' problem of inflation
 \cite{La:1989za} was first obtained in
BD without fine tuning.

On small scales after the introduction of the theory, Brans found four families of static spherically symmetric solutions \cite{Brans:1962zz}. For a long time many authors claimed that these solutions can describe non-trivial black holes different from those of GR. However, it was proven by Hawking \cite{Hawking:1972qk}, that all those  spherically symmetric black hole solutions are the same as in GR. This result was further extended  to scalar-tensor theories \cite{Sotiriou:2011dz} and to compact objects in the presence of a cosmological constant \cite{Bhattacharya:2015iha}. In \cite{Campanelli:1993sm} it was claimed that black hole solutions in the BD theory were found violating the WEC. However,
in \cite{Vanzo:2012zu} it was  showed that their solutions describe either wormholes or naked singularities. It was further proved in \cite{Agnese:1995kd} that  the static spherically symmetric solutions of BD theory describe either wormholes or naked singularities. A more extensive study was performed in \cite{Faraoni:2016ozb} where it was shown that, the static and  spherically symmetric BD solutions of scalar-tensor gravity, analyzed in both the Jordan and the Einstein conformal frames, describe wormholes, naked singularities or the Schwarzschild solution. Thus, they do not describe black hole solutions besides those in GR.

If an electromagnetic field is introduced then in four dimensions the local solutions
reduce to the Reissner-Nordstr\"{o}m black hole solution with a constant
scalar field, as it  was proven in \cite{Hawking:1972qk} using the WEC.
However, in higher dimensions the vacuum
Brans-Dicke-Maxwell theory has black hole solutions \cite{Cai:1996pj}. This is a consequence of the presence
of the electromagnetic field in the scalar field equation and in this way it can be considered as a
source of a non trivial  scalar field.

Recently a modification of BD gravity theory was proposed \cite{Kofinas:2015nwa} in which the scalar field, except its
coupling to the metric it is also coupled to matter \cite{Kofinas:2015sjz}. This coupling, except the BD parameter
 $\omega$,  is introducing another parameter in the coupling of the scalar field to gravity. This coupling is   a new scale in the theory and modifies the matter content of the BD theory, and this scale appears in the vacuum equations of the modified BD theory. In \cite{Kofinas:2016fcp}   the cosmological implications of an extended
BD theory presented in \cite{Kofinas:2015sjz} was discussed. The  new mass scale introduced  in the theory  modifies the  Friedmann equations
with field-dependent corrected kinetic terms. In the general solutions of a radiation
universe  it was found that there are branches
with complete removal of the initial singularity,  while at the
same time a transient accelerating period can occur within
deceleration. Entropy production is also possible in the early universe.
In the dust era, late-times acceleration has been
found numerically in agreement with the correct behaviour of the
density parameters and the dark energy equation of state, while
the gravitational constant has only a slight variation over a
large redshift interval in agreement with observational bounds.

Motivated by the cosmological results of introducing another  coupling of the scalar field to gravity in the BD theory,
in this work we will investigate what are the effects of the new coupling parameter  on small scales. As we already discussed,
in BD theory all the static spherically symmetric solutions of the theory, except the Schwarzschild solution, they  describe either wormholes or naked singularities \cite{Lobo:2010sb}. To evade this problem on has to introduce a potential and in this case non-trivial black hole solutions can be obtained \cite{Campanelli:1993sm,Gao:2006dx}. Therefore, it is interesting to see if the introduction of the new coupling parameter which it does not introduce
new extra matter in the BD theory in the form of a potential and it cannot be absorbed in the redefinition of $\omega$, leads to new spherically symmetric  black hole or wormhole solutions.

  For this, after reviewing in Section \ref{Sec2}  the BD vacuum solutions, we solve the Einstein and scalar equations of this modified BD gravity theory in Section \ref{Sec3}. In this Section we perform a detailed investigation of the solutions. By varying this new coupling
we find two branches of solutions which describe naked singularities and they generate new wormhole geometries whose behaviour depends on the coupling of the scalar field to matter. We did not find new black hole solutions, therefore our solutions are in accordance with the two previously stated theorems  \cite{Hawking:1972qk,Agnese:1995kd}. As it is well known, the wormhole solutions violate the WECs. For this reason we also study the WEC and we find that they are violated. Finally, in Section \ref{Sec5} are our conclusions.

\section{Brans-Dicke vacuum solutions }
\label{Sec2}

In this section we will review  the local solutions of the BD theory
 mainly following the work of \cite{Cai:1996pj}.
 In the $D(\ge 4$) dimensions, the action of the vacuum
BD theory is given by
\begin{equation}
I=\frac{1}{16\pi}\int d^Dx\sqrt{-g}(\phi R-\frac{\omega}{\phi}
g^{\mu\nu}\nabla_{\mu}\phi\nabla_\nu\phi
)~. \label{BDF}
\end{equation}
 In this Jordan frame version of BD
theory, test particles have constant rest
 mass and
move along the geodesics. That is, matter fields are coupled to
gravity
 only via the metric. Varying (\ref{BDF})
yields the equations of motion,
\begin{eqnarray}
\phi G_{\mu\nu}&\equiv & \phi (R_{\mu\nu}-\frac{1}{2}g_{\mu\nu}R)
\nonumber\\
 &=&\frac{\omega}{\phi}\left [\nabla _{\mu}\phi \nabla _{\nu}\phi -
\frac{1}{2}g_{\mu\nu}(\nabla \phi )^2\right ] \nonumber\\
\nonumber\\
&+&\nabla _{\mu}\nabla _{\nu}\phi -g_{\mu\nu}\nabla ^2 \phi~,\label{EinsEg} \\
\nabla ^2 \phi &=&0~,
\label{KGEg}
\end{eqnarray}
where $d=D-3$. Solving (\ref{EinsEg})-(\ref{KGEg}) we consider the following conformal transformation
\begin{equation}
g_{\mu\nu}=\Omega ^2 \bar{g}_{\mu\nu}~, \label{conf1}
\end{equation}
 with
\begin{equation}
\Omega ^{-(d+1)}=\phi~, \label{conf2}
\end{equation}
and
\begin{equation}
\bar{\phi}=\sqrt{2a}\int ^{\phi} \frac{d\phi}{\phi}=
\sqrt{2a}\ln\phi,~~~ a=\frac{d+2}{d+1}+\omega~, \label{transphi}
\end{equation}
the BD theory (\ref{BDF}) can be transformed into the Einstein frame version of BD
theory with a minimally coupled scalar field
$(\bar{\phi})$
\begin{equation}
\bar{I}=\frac{1}{16\pi}\int
d^Dx\sqrt{-\bar{g}}[\bar{R}-\frac{1}{2} (\bar{\nabla}\bar{\phi}
)^2 ]~,\label{tranfFBD}
\end{equation}
 where $\bar{R}$ and $\bar{\nabla}$ are the scalar curvature and covariant
 differentiation in the new metric $\bar{g}_{\mu\nu}$, respectively.
We note

\begin{itemize}

\item
 Relation  (\ref{transphi}) implies
 $a>0$ ($ \omega >-\frac{d+2}{d+1}$), and one has
$\bar{\phi}=0$ at spacelike infinity.

\item The action under the conformal
transformation gets a simpler form of a minimally coupled scalar field.

\item The Brans-Dicke theory (\ref{BDF}) is
 equivalent to the theory (\ref{tranfFBD}) up to a conformal transformation. However, note that in the
 Einstein frame,
 a test particle will take variable rest mass with
spacetime and is no longer going to move along the geodesics. This physical inequivalence can be understood from the conformal transformation
of the metric (\ref{conf1}) and (\ref{conf2}). The conformal tranformation depends on the scalar field $\phi$ which parametrizes  the matter of the theory. Therefore the physical behavior of the theory can only be understood if the coupling to matter is specified.

Hence, one can argue that the two theories are equivalent from a mathematical point of view but not from a physical one.

\end{itemize}

Varying the action (\ref{tranfFBD})  we can obtain the equations of
motion which are connected with the equations of motion of  (\ref{EinsEg})-(\ref{KGEg})
through the relation
\begin{equation}
 (g_{\mu\nu}, \phi )=(e^{-\frac{2}{(d+1)\sqrt{2a}}
 \bar{\phi}}
\bar{g}_{\mu\nu}, e^{\frac{1}{\sqrt{2a}}\bar{\phi}} )~. \label{equiv}
\end{equation}
Introducing  isotropic
 coordinates \cite{Xanthopoulos:1989kb}
\begin{eqnarray}
d\bar{s}^2&=&-e^{f}dt^2+e^{-h}(d\rho^2 +\rho^2d\Omega ^2_{d+1})~, \label{BDmetric}
\end{eqnarray}
in the $D$-dimensional vacuum BD theory, using (\ref{equiv}), we
can obtain its solution,
\begin{eqnarray}
ds^2&=&\Omega ^2d\bar{s}^2=\left (\frac{\rho^d+\rho_o^d}{\rho^d-\rho_o^d}
\right )^{\frac{2}{d+1}\left [\frac{(d+1)(1-\gamma ^2)}{ad}
\right ]^{1/2}} d\bar{s}^2~,\label{BDorig}\\
\phi &=&\left (\frac{\rho^d-\rho_o^d}{\rho^d+\rho_o^d}\right )^{\left
[\frac{(d +1)(1-\gamma ^2)}{ad}\right ]^{1/2}}~,
\end{eqnarray}
where $\gamma$ is a constant and  $d\bar{s}^2$ is given by Eq. (\ref{BDmetric}).  It is easy
to show that the
 solution (\ref{BDorig}) is asymptotically flat and the point $\rho=\rho_o$
 corresponds to
 a naked singularity still.  This can be found from calculating
  the scalar curvature of the solution (\ref{BDorig}) through the relation,
\begin{eqnarray}
R&=&\Omega ^{-2}\bar{R}-2(d+2)\Omega ^{-3}\bar{g}^{\mu\nu}
 \bar{\nabla} _{\mu}\bar{\nabla} _{\nu}\Omega    \nonumber\\
&-&(d+2)(d-1)\Omega ^{-4}\bar{g}^{\mu\nu}\bar{\nabla}_{\mu} \Omega
\bar{\nabla} _{\nu}\Omega~,
\end{eqnarray}
and show that it diverges at zero areal radius. Again we observe,

\begin{itemize}

\item When $\gamma=1$, the solution (\ref{BDorig}) is reduced to
the
 D-dimensional Schwarzschild solution with the constant scalar
  field ($\phi =1$). In that
case, the BD theory degenerates into  the Einstein theory of
gravitation.

\item The scalar  $\phi$ in the BD theory belongs to
the
 region $\phi \in (0, 1]$. From the
 action (\ref{tranfFBD}) we can see that the equations of motion
 remain unchanged under
 the transformation: $\bar{\phi}\rightarrow -\bar{\phi}$. Thus, we can
obtain  another  solution of the vacuum BD theory,
\begin{eqnarray}
ds^2&=&\left (\frac{\rho^d-\rho_o^d}{\rho^d+\rho_o^d}\right )^{\frac{2}{d+1}
\left [\frac{(d+1)(1-\gamma ^2)}{ad}\right ]^{1/2}}d\bar{s}^2~,\label{DoubBd} \\
\phi &=&\left (\frac{\rho^d+\rho_o^d}{\rho^d-\rho_o^d}\right )^{\left [\frac{
(d+1)(1 -\gamma ^2)}{ad}\right ]^{1/2}}~,
\end{eqnarray}
where $d\bar{s}^2$ is still given by Eq. (\ref{BDmetric}). In this
case, the scalar
 field $\phi$ takes values in the region $ [1,\infty)$. But the spacetime
is  still asymptotically flat  region and the point $\rho=\rho_o$
is a curvature singularity unless $\gamma=1$. When $\gamma=1$, the
scalar field is a constant and the solution (\ref{DoubBd}) is the
$D$-dimensional
 Schwarzschild solution. We note that $ \phi=0$ corresponds to  infinite gravitational coupling while $\phi=\infty$ corresponds to zero gravitational coupling, so $\phi$ should be allowed to diverge or vanish only at singularities.

\end{itemize}

The metric \eqref{BDorig} is actually Brans Class 1 solution since if we substitute $D=4$, then $d=1$ and $\alpha=\frac{3+2\omega}{2}$, which is related with our $\lambda$ parameter by $\frac{1}{\alpha}=\frac{1}{\omega+3/2}=\lambda$ (see Section \ref{Sec3}). We substitute all these in the line element, together with the form of $d\bar{s}^2$ and we obtain
 \begin{align}
     ds^2= -\left(\frac{1-\rho_o/\rho}{1+\rho_o/\rho}\right)^{\alpha(\lambda,\gamma)+2\gamma}dt^2+\left(1-\frac{\rho_o^2}{\rho^2}\right)^2\left(\frac{1+\rho_o/\rho}{1-\rho_o/\rho}\right)^{\alpha(\lambda,\gamma)+2\gamma} \left[ d\rho^2+\rho^2 d\Omega^2\right]
 \end{align}
 which is exactly what we find in \eqref{e>0,metric limit to BD}.
 The requirement that the scalar field $\bar{\phi}$, in the Einstein frame, is real implies $\alpha>0\,(\omega>-\frac{3}{2} \text{ in D=4})$ hence, only Brans Class I solution can be obtained in the Jordan frame. However, there are other three classes of Brans solutions which correspond to $\omega<-\frac{3}{2}$ (for a detailed discussion on the  Brans solutions see the review \cite{Faraoni:2016ozb}.


Therefore,  again  the  black hole solution of the
 vacuum BD theory is the Schwarzschild solution with a constant scalar
 field.

\section{Modified Brans-Dicke theory}
\label{Sec3}

We will consider a modified BD theory presented in \cite{Kofinas:2015nwa} and
described by the following equations
\begin{eqnarray}
&&\!\!\!\!\!\!\!G^{\mu}_{\,\,\,\nu}\!=\!\frac{8\pi}{\phi}
(T^{\mu}_{\,\,\,\nu}+\mathcal{T}^{\mu}_{\,\,\,\,\nu})~,
\label{elp}\\
&&\!\!\!\!\!\!\!T^{\mu}_{\,\,\,\nu}\!=\!\frac{\phi}{2\lambda(\nu\!+\!8\pi\phi^{2})^{2}}\Big{\{}
2\big[(1\!+\!\lambda)\nu\!+\!4\pi(2\!-\!3\lambda)\phi^{2}\big]\phi^{;\mu}\phi_{;\nu}
\!-\!\big[(1\!+\!2\lambda)\nu\!+\!4\pi(2\!-\!3\lambda)\phi^{2}\big]\delta^{\mu}_{\,\,\,\nu}
\phi^{;\rho}\phi_{;\rho} \Big{\}}
\!+\!\frac{\phi^{2}}{\nu\!+\!8\pi\phi^{2}}
\big(\phi^{;\mu}_{\,\,\,\,;\nu}\!-\!\delta^{\mu}_{\,\,\,\nu}\Box\phi\big)~,
\nn\\
\label{utd}\\
&&\!\!\!\!\!\!\!\Box\phi\!=\!4\pi\lambda\mathcal{T}~,\label{lrs}\\
&&\!\!\!\!\!\!\!\mathcal{T}^{\mu}_{\,\,\,\,\nu;\mu}\!=\!\frac{\nu}{\phi(\nu\!+\!8\pi\phi^{2})}
\mathcal{T}^{\mu}_{\,\,\,\,\nu}\phi_{;\mu}\,,
\label{idj}
\end{eqnarray}
where $\mathcal{T}^{\mu}_{\,\,\,\,\nu}$ is the energy momentum tensor of matter (e.g. the standard matter-radiation in the case of cosmology).
So we can see that matter couples directly to the scalar field $\phi$. Since in the action of BD
gravity the scalar field is non-minimally coupled to the curvature, the same mechanism could also lead to a coupling
between the scalar and matter fields, as happens here. Various studies have analyzed the exchange of energy from
ordered motion by entropy generation due to bulk viscosity. Of course, the parameters $\lambda, \nu$ should be such that the equivalence principle is not violated at the ranges
that it has been tested. However, the new parameter $\nu$ violates the exact conservation of the matter
energy-momentum tensor $\mathcal{T}^{\mu}_{\,\,\,\,\nu}$ in (\ref{idj}). Interesting cosmological results were obtained in \cite{Kofinas:2015sjz} as we have already discussed.
  For $\nu=0$ the system
(\ref{elp})-(\ref{idj}) reduces to the BD equations of motion (with unit velocity
of light)
\begin{eqnarray}
G^{\mu}_{\,\,\,\nu}\!\!&=&\!\!\frac{8\pi}{\phi}(T^{\mu}_{\,\,\,\nu}+\mathcal{T}^{\mu}_{\,\,\,\,\nu})~,
\label{kns}\\
T^{\mu}_{\,\,\,\nu}\!\!&=&\!\!\frac{2-3\lambda}{16\pi\lambda\phi}\Big(
\phi^{;\mu}\phi_{;\nu}\!-\!\frac{1}{2}\delta^{\mu}_{\,\,\,\nu}
\phi^{;\rho}\phi_{;\rho} \Big)
\!+\!\frac{1}{8\pi}\big(\phi^{;\mu}_{\,\,\,\,;\nu}\!-\!\delta^{\mu}_{\,\,\,\nu}\Box\phi\big)~,
\label{qqd}\\
\Box\phi\!\!&=&\!\!4\pi\lambda\mathcal{T}~,\label{lwn}\\
\,\,\mathcal{T}^{\mu}_{\,\,\,\,\nu;\mu}\!\!&=&\!\!0\,,
\label{jrk}
\end{eqnarray}
which is described by the action
\begin{equation}
S_{BD}=\frac{1}{16\pi}\int \!d^{4}x \,\sqrt{-g} \,\Big(\phi R-\frac{\omega_{BD}}{\phi}
g^{\mu\nu}\phi_{,\mu}\phi_{,\nu}\Big)+\int \!d^{4}x \,\sqrt{-g} \,L_{m}\,,
\label{jsw}
\end{equation}
where $L_{m}(g_{\kappa\lambda},\Psi)$ is the matter Lagrangian
depending on some extra fields $\Psi$.
The parameter $\lambda\neq 0$ is related to the standard BD parameter $\omega_{BD}=
\frac{2-3\lambda}{2\lambda}$.

As we discussed in the Introduction in this work we are interested to see what is the effect of the new
 coupling $\nu$ on local spherically symmetric solutions of the vacuum BD theory,  without extra matter. Therefore we will study the
   modified BD theory  with $\mathcal{T}^{\mu}_{\,\,\,\,\nu}=0$.
Although the extra matter vanishes, it leaves an impact on the vacuum equation (\ref{elp})
through the parameter $\nu$, and this is the novel difference compared to the vacuum BD
equation (\ref{kns}). This vacuum theory arises from the action \cite{Kofinas:2015sjz}
\begin{equation}
S=\frac{\eta}{2(8\pi)^{3/2}}\int\!d^{4}x\,\sqrt{-g}\,
\Big[\sqrt{|\nu\!+\!8\pi\phi^{2}|}\,R-\frac{8\pi}{\lambda}\,
\frac{\nu\!+\!4\pi(2\!-\!3\lambda)\phi^{2}}{|\nu\!+\!8\pi\phi^{2}|^{3/2}}
g^{\mu\nu}\phi_{,\mu}\phi_{,\nu}\Big]\,,
\label{jgk}
\end{equation}
where $\eta=\text{sgn}(\phi)$. It is clear in the above action that the kinetic term of the scalar field is modified compared to the original  Brans-Dicke theory and it results to a modified energy momentum tensor for the scalar field.

In this work we will consider the modified BD theory given by the action (\ref{jgk}) and study spherically symmetric solutions of this theory.
It  is useful for the following analysis of spherically symmetric
solutions to transform the action (\ref{jgk}) to its canonical form. Following the discussion in the previous Section, consider the conformal transformation
\begin{equation}
\tilde{g}_{\mu\nu}=\Omega^{2}(\phi)g_{\mu\nu}\,,
\label{dlt}
\end{equation}
where
\begin{equation}
\Omega=\Big(\frac{|\nu\!+\!8\pi\phi^{2}|}{8\pi}\Big)^{\frac{1}{4}}\,,
\label{mnb}
\end{equation}
together with a field redefinition from the field $\phi(x)$ to the new field $\sigma(x)$
defined by
\begin{equation}
\frac{d\phi}{d\sigma}=\sqrt{\frac{|\lambda|}{16\pi}}\,\sqrt{|\nu\!+\!8\pi\phi^{2}|}\,.
\label{jcy}
\end{equation}
The action (\ref{jgk}) takes the form
\begin{equation}
S=\frac{\eta}{16\pi}\int\!d^{4}x\,\sqrt{-\tilde{g}}\,\Big(\tilde{R}-
\frac{1}{2}\epsilon\epsilon_{\lambda}\tilde{g}^{\mu\nu}\sigma_{,\mu}\sigma_{,\nu}\Big)\,,
\label{vtk}
\end{equation}
where $\epsilon=\text{sgn}(\nu\!+\!8\pi\phi^{2})$, $\epsilon_{\lambda}=\text{sgn}(\lambda)$.
The Lagrangian (\ref{vtk}) refers to the Einstein frame where the gravitational coupling is a
true constant and the field $\sigma$ behaves as a usual scalar field. In this work we are looking for black hole and wormhole solutions.
The action (\ref{vtk})  describes in the Einstein frame the kinetic energy of a scalar field coupled to gravity. We will look first for
spherically symmetric black hole solutions in this frame. For this reason we will assume  $\epsilon\epsilon_{\lambda}>0$ for the scalar field to have a physically propagating   mode. This condition is a necessary condition in order not to have instabilities in our theory. However, as we will see in the Jordan frame we will find wormhole solutions in which the energy conditions are violated \cite{Rubakov:2014jja} and ghost-like instabilities appear. Therefore, this condition does not guarantee the stability of the theory under perturbations. This behavior is not a surprise since stable wormholes can only be supported in higher-derivative theories
of the beyond-Horndeski type \citep{Evseev:2017jek,Franciolini:2018aad,Mironov:2018uou}, a class in which our theory \eqref{jgk} clearly does not belong to.

We assume throughout that $\epsilon\epsilon_{\lambda}=1$.
For $\epsilon>0$, the integration of equation (\ref{jcy}) gives
\begin{equation}
\sigma=
\sqrt{\frac{2}{|\lambda|}}\,\ln{\Big|4\pi\phi\!+\!\sqrt{2\pi}\sqrt{\nu\!+\!8\pi\phi^{2}}\Big|}\,,
\label{kil}
\end{equation}
where an additive integration constant $\sigma_{0}$ has been absorbed into $\sigma$. Inversely,
\begin{equation}
\phi=\frac{s}{8\pi}\Big(e^{\sqrt{\frac{|\lambda|}{2}}\,\sigma}\!-\!2\pi\nu
e^{-\sqrt{\frac{|\lambda|}{2}}\,\sigma}\Big)\,,
\label{nui}
\end{equation}
where $s=\text{sgn}(4\pi\phi\!+\!\sqrt{2\pi}\sqrt{\nu+8\pi\phi^{2}})=
\text{sgn}\Big(e^{\sqrt{\frac{|\lambda|}{2}}\,\sigma}\!+\!2\pi\nu
e^{-\sqrt{\frac{|\lambda|}{2}}\,\sigma}\Big)$. The conformal factor $\Omega$ in terms
of the new field $\sigma$ takes the form
\begin{equation}
\Omega=\frac{1}{\sqrt{8\pi}}\Big|e^{\sqrt{\frac{|\lambda|}{2}}\,\sigma}\!+\!2\pi\nu
e^{-\sqrt{\frac{|\lambda|}{2}}\,\sigma}\Big|^{\frac{1}{2}}\,.
\label{jry}
\end{equation}
For the physically more interesting case with $\phi>0$, the absolute value in (\ref{jry})
disappears.

For $\epsilon<0$, the integration of equation (\ref{jcy}) gives
\begin{equation}
\sigma=\sqrt{\frac{2}{|\lambda|}}\,\arcsin{\Big(\sqrt{\frac{8\pi}{|\nu|}}\,\phi\Big)}\,,
\label{pou}
\end{equation}
where again an additive integration constant $\sigma_{0}$ has been absorbed into $\sigma$ and
it is $-\frac{\pi}{2}<\sqrt{\frac{|\lambda|}{2}}\,\sigma<\frac{\pi}{2}$.
Inversely,
\begin{equation}
\phi=\sqrt{\frac{|\nu|}{8\pi}}\,\sin\Big(\sqrt{\frac{|\lambda|}{2}}\,\sigma\Big)\,.
\label{yun}
\end{equation}
The conformal factor $\Omega$ in terms of the new field $\sigma$ takes the form
\begin{equation}
\Omega=\Big(\frac{|\nu|}{8\pi}\Big)^{\frac{1}{4}}\,
\Big[\cos\Big(\sqrt{\frac{|\lambda|}{2}}\,\sigma\Big)\Big]^{\frac{1}{2}}\,.
\label{dfr}
\end{equation}

After the solution of the fields $\tilde{g}_{\mu\nu},\sigma$ governed by the action (\ref{vtk})
has been derived, the solution for the initial fields $g_{\mu\nu},\phi$ is found through the
equations (\ref{dlt}), (\ref{nui}), (\ref{yun}) as functions of $\sigma$. The action (\ref{vtk})
defines Einstein gravity minimally coupled to a scalar field whose equations of motion are
\begin{eqnarray}
&&\tilde{G}_{\mu\nu}=\frac{1}{2}\sigma_{,\mu}\sigma_{,\nu}-\frac{1}{4}
\tilde{g}_{\mu\nu}\tilde{g}^{\kappa\lambda}\sigma_{,\kappa}\sigma_{,\lambda}~,\label{ekf}\\
&&\tilde{\Box}\sigma=0\label{wkf}\,.
\end{eqnarray}
The solution of this system of a minimally coupled scalar field to gravity,
assuming spherical symmetry, has been found in \cite{Xanthopoulos:1989kb}.
In the Einstein frame, we consider a static spherically symmetric line element in isotropic
coordinates
\begin{equation}
d\tilde{s}^{2}=-e^{f}dt^{2}+e^{-h}\big[d\rho^{2}+\rho^{2}\big(d\theta^{2}+\sin^{2}\!\theta
\,d\varphi^{2}\big)\big]\,,
\label{kwff}
\end{equation}
where $f,h$ are functions of the radial coordinate $\rho$ (we keep the symbol $r$ for the radius
in the standard coordinates). Due to the symmetry it is also $\sigma(\rho)$.
The solution of the system (\ref{ekf}), (\ref{wkf}) is the following \cite{Xanthopoulos:1989kb}
\begin{eqnarray}
&&\sigma=2\sqrt{1\!-\!\gamma^{2}}\,\,\ln\frac{\rho-\rho_{o}}{\rho+\rho_{o}}\label{ied}\\
&&e^{f}=\Big(\frac{\rho-\rho_{o}}{\rho+\rho_{o}}\Big)^{\!2\gamma}\label{kwf}\\
&&e^{-h}=\Big(1-\frac{\rho_{o}^{2}}{\rho^{2}}\Big)^{\!2}\,
\Big(\frac{\rho+\rho_{o}}{\rho-\rho_{o}}\Big)^{\!2\gamma}\label{kwf2}\,,
\end{eqnarray}
where $\rho_{o}>0,\gamma$ are integration constants. The reason why we restrict $\rho_o$ in this range of values will become clear in the next sections, where the limit to the Schwarzschild solution will be considered.

Let us express the quantity $\Omega$ as a function of $\rho$ through equation (\ref{ied}).
For $\epsilon>0$ it is
\begin{equation}
\Omega=\frac{1}{\sqrt{8\pi}}\Bigg{|}\Big(\frac{\rho-\rho_{o}}{\rho+\rho_{o}}\Big)
^{\!\!\sqrt{2|\lambda|(1-\gamma^{2})}}+2\pi\nu\Big(\frac{\rho+\rho_{o}}{\rho-\rho_{o}}\Big)
^{\!\!\sqrt{2|\lambda|(1-\gamma^{2})}}\Bigg{|}^{\frac{1}{2}}\,,
\label{jer}\end{equation}
while for $\epsilon<0$,
\begin{equation}
\Omega=\Big(\frac{|\nu|}{8\pi}\Big)^{\frac{1}{4}}\,
\Big[\cos\Big(\sqrt{2|\lambda|(1\!-\!\gamma^{2})}\,\,\ln\frac{\rho-\rho_{o}}{\rho+\rho_{o}}\Big)\Big]
^{\frac{1}{2}}\,.
\label{djr}\end{equation}
Note that in the limit $\rho\rightarrow\rho_{o}$, it is $\Omega\rightarrow\infty$ for $\epsilon>0$,
while $\Omega$ remains finite for $\epsilon<0$.

Going back to the original frame we have in the isotropic coordinates
\begin{equation}
ds^{2}=-\Omega^{-2}e^{f}dt^{2}+\Omega^{-2}e^{-h}
\big[d\rho^{2}+\rho^{2}\big(d\theta^{2}+\sin^{2}\!\theta\,d\varphi^{2}\big)\big]\,.
\label{edg}
\end{equation}
The metric in the standard coordinates is obtained setting
\begin{equation}
r=\rho\Omega^{-1} e^{-\frac{h}{2}}\,,\label{iwf}
\end{equation}
and then
\begin{equation}
ds^{2}=-\Omega^{-2}e^{f}dt^{2}+\frac{r^{2}}{\rho^{2}}\frac{dr^{2}}{(\frac{dr}{d\rho})^{2}}
+r^{2}(d\theta^{2}+\sin^{2}\!\theta\, d\varphi^{2})\,.
\label{erw}
\end{equation}
Let us now take a closer look at the behaviour of this metric for both branches.

\subsection{ Spherically symmetric solutions in the Jordan frame}
\label{Sec4}

 We consider the line element in the Jordan frame
\begin{align}
ds^2=-\Om^{-2}e^f dt^2+\Om^{-2}e^{-h}(d\rho^2+\rho^2d\Om^2)~,\label{line-elem}
\end{align}
in isotropic coordinates $(t,\rho,\theta,\phi)$, where $d\Omega^2=d\theta^2+\sin^2\theta d\phi^2$ is the line element of the unit 2-sphere.
The exponentials are given by \eqref{kwf},\,\eqref{kwf2} and the conformal factors by \eqref{jer} and \eqref{djr} for $\epsilon>0$ and $\epsilon<0$, respectively.
It must be $\rho>\rho_o>0$ and $0\leq \gamma^2 \leq 1\nonumber$.
The $\lambda\neq 0$ parameter is related to the standard BD parameter by $\lambda=\frac{1}{\omega+3/2}$. Hence in our model, $\lambda$ and $\nu$ are parameters of the theory and $\rho_o$ and $\gamma$ are the parameters of this specific family of solutions.

\subsection{Branch $\epsilon<0$}

As previously stated, we require $\epsilon<0$ and $\epsilon_\lambda<0$ to avoid ghost solutions of the field $\sigma$. That is
\begin{align*}
    &\epsilon=\text{sgn}(\nu+8\pi\phi^2)<0\,,\\
    &\epsilon_\lambda=\text{sgn}(\lambda)<0\,.
\end{align*}
Substituting \eqref{ied} into \eqref{yun} and using the preceding inequalities we can find that when $\nu<0$ and  $\rho>\rho_o\frac{e^K+1}{e^K-1}$, where $K=\frac{\pi}{2\sqrt{2|\lambda|(1-\gamma^2)}}$, the relation $\epsilon<0$ is always satisfied. Basically, this is the range of $\rho$ such that $-\frac{\pi}{2}<\sqrt{\frac{|\lambda|}{2}\sigma}<\frac{\pi}{2}$. Therefore, in the general case, we will only consider such values of $\rho$ in the following analysis. Moreover, the condition $\epsilon_\lambda<0$ is the equivalent to $\omega<-3/2$.
\subsubsection{Scalar field}
The Brans-Dicke scalar field is given by the relation
\begin{align}
    \phi=\sqrt{\frac{|\nu|}{8\pi}}\sin{\left(\sqrt{2 |\lambda|(1-\gamma^2)}\ln \frac{\rho-\rho_o}{\rho+\rho_o} \right)}\,.\label{BDscalar-e<0}
\end{align}
Note that $\phi$ becomes constant when $\gamma=\pm1$, and the theory reduces to GR.
Moreover, the scalar field vanishes in the limits $\rho\rightarrow\infty$ or $\lambda\rightarrow0 ~(i.e.\,\,  \omega\rightarrow\infty)$, which means the effective gravitational constant diverges.
\subsubsection{Metric Components}
The metric components in isotropic coordinates are given by
\begin{align*}
g_{tt}
=&-\Big(\frac{8\pi}{|\nu|}\Big)^{\frac{1}{2}}\sec\Big(\sqrt{2 |\lambda|(1-\gamma^2)}\ln \frac{\rho-\rho_o}{\rho+\rho_o} \Big)\Big(\frac{\rho-\rho_{o}}{\rho+\rho_{o}}\Big)^{2\gamma}~,\\[0.2cm]
g_{\rho\rho}
=&\Big(\frac{8\pi}{|\nu|}\Big)^{\frac{1}{2}}\sec\Big(\sqrt{2 |\lambda|(1-\gamma^2)}\ln \frac{\rho-\rho_o}{\rho+\rho_o} \Big)\left(1-\frac{\rho_o^2}{\rho^2}\right)^2\left(\frac{1+\rho_o/\rho}{1-\rho_o/\rho}\right)^{2\gamma}\,.
\end{align*}
The appearance of the secant function prevents $g_{tt}$ and $g_{\rho\rho}$ from vanishing, however it causes divergences at the points that satisfy the equation $\sqrt{2 |\lambda|(1-\gamma^2)}\ln \frac{\rho-\rho_o}{\rho+\rho_o}=-\frac{\pi}{2}$ i.e. $\rho=\rho_o\frac{e^K+1}{e^K-1}$, if $\gamma\neq\pm1$.

It is rather easy to check the asymptotic behavior of our metric. A straightforward calculation gives
\begin{align}
\lim_{\rho\rightarrow\infty}g_{tt}=-\sqrt{\frac{8\pi}{\abs{\n}}}\quad ,\quad
\lim_{p\rightarrow\infty}g_{\rho\rho}=\sqrt{\frac{8\pi}{\abs{\n}}}\,.\label{eq:asymtlimits,e<0}
\end{align}
The fact that $\nu$ is a parameter of a theory and not a dynamic variable, means that we can absorb the above factors in the line element by  redefining $dt$ and $d\rho$. Thus, in this case the line element (\ref{line-elem}) with $\Omega$ given by (\ref{djr}), describes an asymptotically flat spacetime. Moreover, note that when $\nu=0$ the line element diverges and the solution does not produce the corresponding solutions of BD. This is not surprising since setting $\nu=0$ automatically corresponds to $\epsilon=\text{sgn}(8\pi\phi^2)>0$ and that is why only the branch $\epsilon>0$ has a correct limit to Brans-Dicke.

Let us now consider two particular cases:

    If $\gamma=1$  then
    \begin{align}
        g_{tt}&=-\Big(\frac{8\pi}{|\nu|}\Big)^{\frac{1}{2}}\left(\frac{\rho-\rho_o}{\rho+\rho_o}\right)^{2}~,\label{g00-e<0-g=1}\\[0.2cm]
        g_{\rho\rho}&=\Big(\frac{8\pi}{|\nu|}\Big)^{\frac{1}{2}}\left( 1+\frac{\rho_o}{\rho}\right)^{4}\,.
    \end{align}
    Now the range $\rho_o<\rho<\rho_o\frac{e^K+1}{e^K-1}$ is physically meanigful.
    Note that $g_{tt}\xrightarrow{\rho\rightarrow\rho_o}0$ while it remains negative for $\rho\neq\rho_o$.
    The BD scalar becomes constant for $\gamma=1$ and as can be seen from the forms of $g_{tt}$ and $g_{\rho\rho}$, the solution reduces to the standard Schwarzschild metric in isotropic coordinates, with mass $M=2\rho_o$.
    In the case $\gamma=-1$ one can find that
    \begin{align}
        g_{tt}=-\Big(\frac{8\pi}{|\nu|}\Big)^{\frac{1}{2}}\left(\frac{\rho+\rho_o}{\rho-\rho_o}\right)^{2}\xrightarrow{\rho\rightarrow\rho_o} -\infty
    \end{align}
    and
    \begin{align}
        g_{\rho\rho}=\Big(\frac{8\pi}{|\nu|}\Big)^{\frac{1}{2}}\left( 1-\frac{\rho_o}{\rho}\right)^4\xrightarrow{\rho\rightarrow\rho_o}0\,.
    \end{align}
    The scalar field is again constant thus the  solution corresponds to Schwarzschild spacetime with a negative mass $M=-2\rho_o$.

\subsubsection{Areal Radius \& Ricci Scalar}
Now we wish to analyze the behavior of the areal radius since it can give us extra information about the geometry but also, help us to deduce  which ranges of the spatial coordinate $\rho$ are physically meaningful. Basically, what we call areal radius is just the radial coordinate of the spherical (Schwarschild) coordinates and that is why we are going to denote it  by $r$. On the other hand, the study of scalar quantities helps us to detect real spacetime singularities since they do not depend on our choice of coordinates. They are invariants and thus, if we manage to find a point where a scalar curvature  diverges we know that it will correspond to a true spacetime singularity.

The areal radius is read off the line element (\ref{edg})  $$r=\rho\Omega^{-1}e^{-\frac{h}{2}}\,$$
 and in this case it is
 \begin{align}
r=&\rho\Omega^{-1}e^{\frac{h}{2}}=\Omega^{-1}\frac{1}{\rho}\frac{(\rho+\rho_o)^{\g    +1}}{(\rho-\rho_o)^{\g-1}}\nonumber\\[0.2cm]
 =&\Big(\frac{|\nu|}{8\pi}\Big)^{-1/4}\,
\sec^{\frac{1}{2}}\Big(\alpha(\lambda,\gamma)\ln \frac{\rho-\rho_o}{\rho+\rho_o} \Big) \frac{1}{\rho} \left(\rho+\rho_o\right)^{1+\gamma}\left(\rho-\rho_o\right)^{1-\g}~,\label{arealradius,e<0}
\end{align}
while its derivative is given by
\begin{align}
\frac{dr}{d\rho}=&\sqrt[4]{\frac{2\pi}{\abs{\nu}}}\frac{1}{p^2}\left(\frac{\rho+\rho_o}{\rho-\rho_o}\right)^{\g}\Bigg[\sqrt{2}\left(    \rho^2+\rho_o^2-2\g\rho\rho_o\right)\cos\left(\alpha(\lambda,\g)\ln\frac{\rho-\rho_o}{\rho+\rho_o}\right) +\nonumber\\[0.2cm]
&  \sqrt{2}\alpha(\lambda,\g)\rho\rho_o\sin\left(\alpha(\lambda,\g)\ln\frac{\rho-\rho_o}{\rho+\rho_o}\right)\Bigg]
\sec^{\frac{3}{2}}\left(\alpha(\lambda,\g)\ln\frac{\rho-\rho_o}{\rho+\rho_o}\right)~.\label{Dareal,e<0}
\end{align}
 Furthermore, the Ricci scalar is given by
\begin{align}
\mathcal{R}=&\sqrt{\frac{\abs{\nu}}{2\pi}}(1-\gamma^2)\rho^4\rho^2_o\frac{(\rho-\rho_o)^{2(\gamma-2)}}{(\rho+\rho_o)^{2(\gamma+2)}} \sec{\left(\alpha(\lambda,\gamma)\ln{\frac{\rho-\rho_o}{\rho+\rho_o}}   \right)}\cdot\nonumber\\[0.25cm]
&\cdot\Bigg\{ 4\cos^2\left( \alpha(\lambda,\gamma)\ln{\frac{\rho-\rho_o}{\rho+\rho_o}} \right) +3|\lambda|\Big[ \cos{\left(\alpha(\lambda,\gamma)\ln{\frac{\rho-\rho_o}{\rho+\rho_o}}   \right)}-5 \Big]\Bigg\}~.
\label{Ricci-e<0}
\end{align}
Note that all of these quantities as well as the metric components, are proportional to $\sec{\left(\alpha(\lambda,\gamma)\ln{\frac{\rho-\rho_o}{\rho+\rho_o}} \right)}$ which forces them to diverge near the point $\rho=\rho_o\frac{e^K+1}{e^K-1}$, if $\gamma\neq\pm1$.
Here a few additional points should be stressed:

      If $\g=1$ then
\begin{align} r&=\left(\frac{8\pi}{\abs{\nu}}\right)^{1/4} \frac{(\rho+\rho_o)^2}{\rho}\;,\label{arealradius-e<0-g=1}\\[0.2cm] \frac{dr}{d\rho}&=\left(\frac{8\pi}{\abs{\nu}}\right)^{1/4}\left(1-\frac{\rho_o^2}{\rho^2}\right)\label{Darealradius-e<0-g=1}~, \end{align}

    if $\g=-1$
\begin{align} r&=\left(\frac{8\pi}{\abs{\nu}}\right)^{1/4} \frac{(\rho-\rho_o)^2}{\rho}\;,\label{arealradius-e<0-g=-1}\\[0.2cm] \frac{dr}{d\rho}&=\left(\frac{8\pi}{\abs{\nu}}\right)^{1/4}\left(1-\frac{\rho_o^2}{\rho^2}\right)\label{Darealradius-e<0-g=-1}~,\end{align}
which means that for these two particular cases, the areal radius decreases for $0<\rho<\rho_o$, has an absolute minimum at $\rho=\rho_o$(whose value is $r=\left(\frac{8\pi}{\abs{\nu}}\right)^{1/4}4\rho_o>0$ if $\g=1$, and $r=0$ if $\g=-1$), and increases for $\rho>\rho_o$. Thus, for $\gamma=-1$ the range $0<\rho<\rho_o$ is unphysical.
Moreover, notice that $r\rightarrow+\infty$ in the limits $\rho\rightarrow 0$ and $\rho\rightarrow\infty$, for both cases $\gamma=\pm1$. Therefore the region near $\rho\rightarrow0$ corresponds to a second asymptotically flat region of spacetime.

For these two particular values of $\gamma$ we know that the scalar field becomes trivial and the solution describes a Schwarzschild spacetime.

 If  $\g \neq \pm1$ the areal radius approaches infinity if $\rho\to\infty$ or $\rho\to\rho_o\frac{e^K+1}{e^K-1}$ where $K=\frac{\pi\alpha(\lambda,\g)}{2}$ and has a  point of minimun value which satisfies the equation
\begin{align}\frac{dr}{d\rho}=0\to -\frac{1}{\alpha(\lambda,\g)}\left( \frac{\rho_o}{\rho}+\frac{\rho}{\rho_o}-2\g \right) = \tan\left(\alpha(\lambda,\g)\ln \frac{\rho-\rho_o}{\rho+\rho_o} \right)\,. \label{throat-location}
\end{align} Since $r$ goes to infinity if $\rho\to\infty$ or $\rho=\rho_o\,\frac{e^K+1}{e^K-1}\geq\rho_o$, the range $0\leq\rho\leq\rho_o\,\frac{e^K+1}{e^K-1}$ is unphysical. By taking a closer look at \eqref{arealradius,e<0} it is evident that $r(\rho)$ is always positive in the range $\rho>\rho_o\,\frac{e^K+1}{e^K-1}$ . Such a behaviour indicates a wormhole structure. To justify this last statement let us represent \eqref{line-elem} in the Morris-Thorne form \cite{Thorne:1988}
\begin{align}
    ds^2=-e^{-2\Phi(r)}dt^2+\frac{dr^2}{1-b(r)/r}+r^2(d\theta^2+\sin^2\theta d\phi^2)\,.\label{MorrisThorne-metric}
\end{align}
Following \cite{Thorne:1988,Visser:1995cc}, certain conditions have to be imposed in order for a line element to be considered as a wormhole spacetime, namely:
\begin{enumerate}
    \item $\frac{b(r)}{r}\leq 1$ for every $[r_{th},+\infty)$, where $r_{th}$ is the radius of the throat. This condition ensures that the proper radial distance defined by $l(r)=\pm\int_{r_{th}}^{r}\frac{dr}{1-\frac{b(r)}{r}}$ is finite everywhere in spacetime.
   Note that in the coordinates $(t,l,\theta,\phi)$ the line element \eqref{MorrisThorne-metric} can be written as $$ds^2=-e^{2\Phi(l)}dt^2+dl^2+r^2(l)(d\theta^2+\sin^2\theta d\phi)\,.$$ In this case the throat radius would be given by $r_{th}=\min\{r(l)\}$.
    \item $\frac{b(r_{th})}{r_{th}}=1$ at the throat. This relation comes from requiring the throat to be a stationary point of $r(l)$. Equivalently, one may arrive at this equation by demanding the embedded surface of the wormhole to be vertical at the throat.
    \item $b'(r)<\frac{b(r)}{r}$ which reduces to $b'(r_{th})< 1$ at the throat. This is known as the flare-out condition since it guarantees $r_{th}$ to be a minimum and not any other stationary point.
    \end{enumerate}

Confronting the two metrics we directly see that the radial coordinate is given by the already known relation \eqref{iwf}. The redshift and shape functions are respectively given by

\begin{align}
        \Phi(r)&=\frac{1}{2}\left( \ln{e^f}+\ln{\Omega^{-2}} \right)\,,\label{redshift-function}\\
        b(r)&=r\left[1-\frac{1}{\Omega^{-2}e^{-h}}\left(\frac{dr}{d\rho}\right)^2\right]~,\label{shape-function}
    \end{align}
or in terms of the coordinates
\begin{align}
    \Phi(r)&=\frac{1}{4}\ln{\left(\frac{8\pi}{|\nu|}\right)}+\frac{1}{2}\Bigg\{ \ln{\left(\frac{\rho(r)-\rho_o}{\rho(r)+\rho_o}\right)^{2\gamma}}-\ln{\left[ \cos\left( \alpha(\lambda,\gamma)\ln{\frac{\rho(r)-\rho_o}{\rho(r)+\rho_o}}  \right)  \right] }   \Bigg\}\;,\label{redshift-e<0}\\[0.2cm]
    \frac{b(r)}{r}&=1-\Bigg\{\frac{\rho(r)^2+\rho_o^2-2\gamma\rho(r)\rho_o+\rho(r)\rho_o\alpha(\lambda,\gamma)\tan\left(\alpha(\lambda,\gamma)\ln{\frac{\rho(r)-\rho_o}{\rho(r)+\rho_o}}\right)}{\sqrt{2}\,(\rho(r)^2-\rho^2_o)}   \Bigg\}^2\;.\label{shape-e<0}
\end{align}
All the afforementioned conditions are verified in the range $[r_{th},+\infty)$.
The throat condition $\frac{b(r_{th})}{r_{th}}=1$ has the same roots with \eqref{throat-location} i.e. it is satisfied when $r$ reaches its minimum value.
 Substituting these roots back in r, one can determine the area of this spatially finite ``bridge'' which is given by $A(\rho)=4\pi r(\rho)^2$ \cite{Visser:1995cc}. Moreover, since from \eqref{arealradius,e<0} we observe that $r\propto|\nu|^{-1/4}$ then, we conclude for the size of the bridge that \begin{equation}
    A(\rho)\propto|\nu|^{-1/2}\,.
\end{equation}
 The Ricci scalar is finite at the throat hence, the solution indeed describes a geometrically traversable wormhole. On the other hand, it diverges as one approaches the second asymptotic spatial infinity  $\rho=\rho_o\,\frac{e^K+1}{e^K-1}$ ($r\rightarrow\infty$) associated with our wormhole. Thus the region near $\rho=\rho_o\,\frac{e^K+1}{e^K-1}$ is asymptotically large in the sense that the proper area of a circle at radius $\rho$, $A(\rho)=4\pi r(\rho)^2 $, goes to infinity as one approaches that region
but not asymptotically flat, making the wormhole asymmetric under the interchange of the two asymptotic regions. This feature is also exhibited in
Brans Class I solution for a specific range of its parameters (see \citep{Visser:1997yn},\cite{Faraoni:2016ozb} for more details).
Also, notice that even though it is not evident from the form of \eqref{shape-e<0}, the parameter $\nu$ determines the behaviour of $b(r)$ as it is hidden inside $\rho(r)$, the inverse of \eqref{arealradius,e<0}. This behaviour is depicted in Fig. \ref{fig:allinplot,e<0}.

 In \cite{Thorne:1988} the authors proved that every wormhole, by definition, must violate the Null Energy Condition (NEC) if one assumes that a perfect fluid generates the wormhole spacetime. This is guaranteed by the flaring-out condition $b'(r)<\frac{b(r)}{r}$. By going in a proper reference frame, i.e. a frame of a static observer with respect to $(r,\theta,\phi)$ coordinates, one can express the energy density $\varrho$ and radial pressure $p_r$ with respect to the isotropic radius by using the known relations \cite{Thorne:1988}
\begin{align}
    \varrho &= \frac{1}{r^2}\,\frac{db}{d\rho}\left(\frac{dr}{d\rho}\right)^{-1}\label{energy-density}\\[0.2cm]
    p_r &= \frac{2}{r}\left(1-\frac{b}{r}\right)\,
    \frac{d\Phi}{d\rho}\left(\frac{dr}{d\rho}\right)^{-1}-\frac{b^3}{r}\,.\label{radial-pressure}
\end{align}
Their explicit forms are given in the appendix \ref{appendixA} and as Fig. \ref{fig:energydensity,e<0} indicates, for general values of the parameters (i.e. for $\gamma\neq\pm1\,,\,\lambda\neq0$ and $\nu\neq0$) the energy density becomes negative close to the throat thus, the WEC is also violated.

\begin{figure}[H]
    \centering
    \includegraphics[scale=0.85]{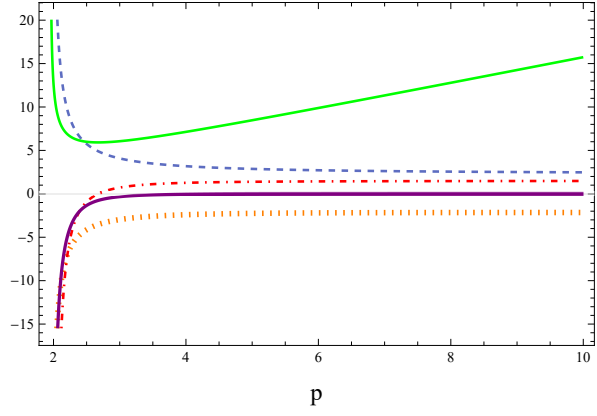}
\caption{Plot of $g_{tt}$(orange-dashed line), $g_{\rho\rho}$(blue-dashed line), areal radius $r$(green solid line), $\frac{dr}{d\rho}$(red dashed-dotted line) and Ricci scalar(purple solid line) versus the isotropic radius with parameter values  $\rho_o=1\,,\gamma=0.2\,,\lambda=-1\,,\nu=3$. All quantities diverge at the point $\rho=\rho_o\,\frac{e^K+1}{e^K-1}$ which, as can be seen, corresponds to $r\rightarrow\infty$. The areal radius (green) has a point of minimum value (vanishing of the red curve) at which $g_{tt}\neq0$ and $g_{\rho\rho},\mathcal{R}$ are finite. This point connects the two asymptotically flat regions at $\rho\rightarrow\infty$ and $\rho\rightarrow\rho_o\,\frac{e^K+1}{e^K-1}$. The scalar field remains finite for every $\rho>\rho_o\,\frac{e^K+1}{e^K-1}$ while it vanishes as $\rho\rightarrow\infty$. }
\label{fig:allinplot,e<0}
\end{figure}
\begin{figure}[H]
    \centering
    \includegraphics[scale=0.85]{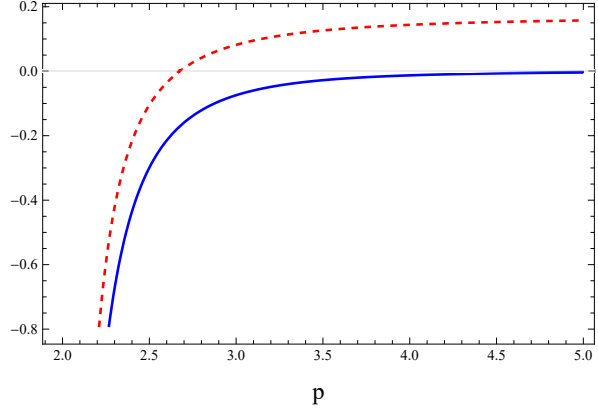}
\caption{Plot of energy density $\varrho$(blue solid curve) and the derivative of $r(\rho)$ with respect to the isotropic radius $\rho$ for $\rho_o=1\,,\gamma=-0.2\,,\lambda=-2\,,\nu=3$. The energy density becomes negative as one approaches the wormhole throat (vanishing of $\frac{dr}{d\rho}$) indicating the presence of  "exotic" matter and the violation of the WEC. }
\label{fig:energydensity,e<0}
\end{figure}
\subsection{Branch $\epsilon>0$}

 We now require $\epsilon>0$ and $\epsilon_\lambda>0$ to avoid ghost solutions of the field $\sigma$. That is
\begin{align*}
    &\epsilon=\text{sgn}(\nu+8\pi\phi^2)>0\,,\\
    &\epsilon_\lambda=\text{sgn}(\lambda)>0\,.
\end{align*}
Following the same procedure as before, we substitute \eqref{ied} into \eqref{nui} and solve the first inequality i.e. $\nu>-8\pi\phi^2$. This leads to
\begin{align}
    \nu\left[ 1-4\pi+4\pi^2\nu\left(\frac{\rho+\rho_o}{\rho-\rho_o}\right)^{2\alpha(\lambda,\gamma)} \right]>-\left(\frac{\rho-\rho_o}{\rho+\rho_o}\right)^{2\alpha(\lambda,\gamma)}\,,
\end{align}
where we have denoted $\alpha(\lambda,\g)\equiv\sqrt{2\abs{\lambda}(1-\g^2)}$ for brevity. It can be seen that $0\leq\alpha(\lambda,\gamma)\leq\sqrt{2\lambda}$.
Numerically it is found that the last relation holds for every $\rho>\rho_o$ and puts no further bounds on the values of $\nu$. Furthermore, the condition $\epsilon_\lambda>0$ corresponds to $\omega>-3/2$.
\subsubsection{Scalar field}
In this case the Brans-Dicke scalar field takes the form
\begin{align}
      &\phi=\frac{s}{8\pi} \bigg| \left( \frac{1-\rho_o/\rho}{1+\rho_o/\rho}\right)^{\alpha(\lambda,\gamma)}-2\pi\n\left( \frac{1+\rho_o/\rho}{1-\rho_o/\rho} \right)^{\alpha(\lambda,\gamma)} \bigg|\label{BDscalar-e>0}\,,\\[0,2cm]
    &s=\text{sgn}(4\pi\phi+\sqrt{2\pi(\nu+8\pi\phi^2)})\,
\end{align}
for which we observe that:
\begin{itemize}
    \item The limit  $\lambda\rightarrow0$ (i.e. $\omega\rightarrow\infty$) corresponds to $\phi\rightarrow\,$constant.
    \item Additionally, for $\gamma=\pm1$ the scalar field also becomes trivial.
    \item When $\gamma\neq\pm1$ and $\nu\neq0$ then $\phi$ diverges in the limit $\rho\rightarrow\rho_o$ thus, the effective gravitational constant vanishes.
    \item In the limit $\nu\rightarrow0$ the scalar field yields
    \begin{align}
         &\phi=\frac{s}{8\pi} \bigg| \left( \frac{1-\rho_o/\rho}{1+\rho_o/\rho}\right)^{\alpha(\lambda,\gamma)} \bigg|\,,\\[0,2cm]
    &s=\text{sgn}(\phi)~.\label{scalar-field,e>0,n=0}
    \end{align}
\end{itemize}
\subsubsection{Metric Components}
The metric components are given by
\begin{align}
g_{tt}=&-8\pi\Bigg| \frac{ \left( \rho^2-\rho_o^2\right)^{\alpha(\lambda,\gamma)}}{(\rho-\rho_o)^{2\alpha(\lambda,\gamma)} +2\pi\nu (\rho+\rho_o)^{2\alpha(\lambda,\gamma)}  }   \Bigg|\Big(\frac{1-\rho_{o}/\rho}{1+\rho_{o}/\rho}\Big)^{2\gamma}\;,\\
g_{\rho\rho}=&\;\,8\pi\Bigg| \frac{ \left( \rho^2-\rho_o^2\right)^{\alpha(\lambda,\gamma)}}{(\rho-\rho_o)^{2\alpha(\lambda,\gamma)} +2\pi\nu (\rho+\rho_o)^{2\alpha(\lambda,\gamma)}  }   \Bigg|\left(1-\frac{\rho_o^2}{\rho^2}\right)^2\left(\frac{1+\rho_o/\rho}{1-\rho_o/\rho}\right)^{2\gamma}
\end{align}
and we calculate
\begin{align}
\lim_{\rho\rightarrow\infty}g_{tt}=-\frac{8\pi}{\abs{2\pi\n+1}}\quad , \quad
\lim_{\rho\rightarrow\infty}g_{\rho\rho}=\frac{8\pi}{\abs{2\pi\n+1}}\,\;.\label{eq:asymtlimits,e>0}
\end{align}
Again, the asymptotic behavior of $g_{tt},g_{\rho\rho}$ depends only on the parameter $\nu$ which means spacetime becomes Minkowski in the large distance limit. We note:

    In the limit $\lambda\rightarrow0\,(\omega\rightarrow\infty)$ the metric takes the form
    \begin{align}
         g_{tt}=-\frac{8\pi}{|1+2\pi\nu|}\left(\frac{1-\rho_o/\rho}{1+\rho_o/\rho}\right)^{2\gamma}\quad,\quad
        g_{\rho\rho}=\frac{8\pi}{|1+2\pi\nu|}\left(1-\frac{\rho_o^2}{\rho^2}\right)^2\left(\frac{1+\rho_o/\rho}{1-\rho_o/\rho}\right)^{2\gamma}\,.
    \end{align}
    This is solution \eqref{ied}-\eqref{kwf2} with the introduction of a new scale, supported by the new parameter $\nu$.

    As $\rho\rightarrow\rho_o$, $g_{\rho\rho}$ vanishes except when $\gamma=1$ for which it blows up.

     If $-1\leq\gamma\leq-\sqrt{\frac{|\lambda|}{|\lambda|+2}}$, the temporal component diverges at $\rho_o$ whereas if $-\sqrt{\frac{|\lambda|}{|\lambda|+2}}\leq\gamma\leq1$ it is $g_{tt}\xrightarrow{\rho\rightarrow\rho_o}0$.

    If $\gamma=1$ then
    \begin{align}
        g_{tt}=-\frac{8\pi}{|1+2\pi\nu|}\left(\frac{1-\rho_o/\rho}{1+\rho_o/\rho}\right)^2\quad,\quad
        g_{\rho\rho}=\frac{8\pi}{|1+2\pi\nu|}\left(1+\frac{\rho_o}{\rho}\right)^4~,
    \end{align} while the Brans-Dicke scalar becomes constant and the solution reduces to the Schwarzschild solution with $M=2\rho_o$.
    On the other hand when $\gamma=-1$ one directly sees that
    \begin{align}
        g_{tt}=-\frac{8\pi}{|1+2\pi\nu|}\left(\frac{1+\rho_o/\rho}{1-\rho_o/\rho}\right)^2\quad,\quad
        g_{\rho\rho}=\frac{8\pi}{|1+2\pi\nu|}\left(1-\frac{\rho_o}{\rho}\right)^4~,
    \end{align}
which corresponds again to the Schwarzschild solution with negative mass $M=-2\rho_o$. Therefore $\rho_o$ can be considered as the mass parameter of the solution. Consequently, we will consider only positive values of $\rho_o$ since, even in the case $\gamma=-1$ we can recover a positive mass by taking $-\rho_o$ to be the mass parameter instead of just $\rho_o$.

     As we consider the limit to Brans-Dicke
    $\nu\rightarrow0$ the metric reads
    \begin{align}
        g_{tt}=-8\pi \left(\frac{1-\rho_o/\rho}{1+\rho_o/\rho}\right)^{2\gamma-\alpha(\lambda,\gamma)}\quad,\quad g_{\rho\rho}=8\pi \left(1-\frac{\rho_o^2}{\rho^2}\right)^2\left(\frac{1+\rho_o/\rho}{1-\rho_o/\rho}\right)^{\alpha(\lambda,\gamma)+2\gamma}\label{e>0,metric limit to BD}
    \end{align} and we notice that
   $2\gamma-\alpha(\lambda,\gamma)>0$ implies $\sqrt{\frac{|\lambda|}{2+|\lambda|}}<\gamma\leq 1\longrightarrow g_{tt}\xrightarrow{\rho\rightarrow\rho_o}0\,$ whereas,
           $2\gamma-\alpha(\lambda,\gamma)<0$ gives $-1\leq\gamma<\sqrt{\frac{|\lambda|}{2+|\lambda|}}\longrightarrow g_{tt}\xrightarrow{\rho\rightarrow\rho_o}\infty$. For the radial component one can calculate that $\alpha(\lambda,\gamma)+2(\gamma-1)>0$ leads to $-1\leq\gamma<\frac{2-|\lambda|}{2+|\lambda|}$ and consequently $g_{\rho\rho}\xrightarrow{\rho\rightarrow\rho_o}0$ while, in the range $\frac{2-|\lambda|}{2+|\lambda|}<\gamma<1$ we obtain $\alpha
    (\lambda,\gamma)+2(\gamma-1)>0$ and it is $g_{\rho\rho}\xrightarrow{\rho\rightarrow\rho_o}\infty$.

    More importantly, solution \eqref{e>0,metric limit to BD} can be identified as Brans Class I solution
    \begin{align}
     ds&^2_{(I)} = -\left(\frac{1-B/\rho}{1+B/\rho}\right)^{2/\lambda_I} dt^2+  \left(1+\frac{B}{\rho}\right)^{4} \left(\frac{1-B/\rho}{1+B/\rho}\right)^{\frac{2(\lambda_I-C-1)}{\lambda_I}}\left(d\rho^2+\rho^2d\Omega_{(2)}^2\right) \label{brans1-metric}\,,\\[0.22cm]
     &\phi_{I} = \phi_0 \left(\frac{1-B/\rho}{1+B/\rho}\right)^{C/\lambda_I}\label{brans1-scalarfield}\;,\\
     &\lambda_I^2=(C+1)^2-C\left(C-\frac{\omega C}{2}\right)~,
    \end{align}
    if we make the substitutions
    \begin{align}
        2\gamma-\alpha(\lambda,\gamma) = \frac{2}{\lambda_I}\;\;,\;\;
        2\gamma+\alpha(\lambda,\gamma) = 2\left(\frac{C+1}{\lambda_I}\right)\;\;,\;\;
        B = \rho_o\;.
    \end{align}
    The above relations yield
    \begin{align}
        \gamma = \frac{C+2}{2\lambda_I}\;,\; \alpha(\lambda,\gamma)=\frac{C}{\lambda_I}\,.\label{CDBtoBDparameters}
    \end{align}
    Thus, just as $(B,C)$  (or $(B,\lambda_I)$ ) are the two independent parameters of Brans Class I solution, in our case they are $(\rho_o,\gamma)$. The Schwarzschild solution is obtained by setting $C=0$, which forces $\lambda_I\pm1$. These values correspond to $\gamma=\pm1$ in our solution which indeed yields the Schwarzschild solution as we saw earlier.

\subsubsection{Areal Radius \& Ricci Scalar}
By substituting the corresponding $\Omega$,  the areal radius and its derivative take the form
\begin{align}
r=&\,\frac{\sqrt{8\pi}}{\rho}\,\Bigg|\frac{(\rho-\rho_o)^{\alpha(\lambda,\gamma)-2(\gamma-1)}\,\cdot\,
(\rho+\rho_o)^{\alpha(\lambda,\gamma)+2(\gamma+1)}}{(\rho-\rho_o)^{2\alpha(\lambda,\gamma)}
+2\pi\nu(\rho+\rho_o)^{2\alpha(\lambda,\gamma)}}\Bigg|^{1/2}\label{arealradius,e>0}~,\\[0.2cm]
\frac{dr}{d\rho}=&\left(\frac{\rho+\rho_o}{\rho-\rho_o}\right)^{\gamma}\frac{1}{\rho^2\Omega}                         \cdot\Bigg\{ (\rho^2+\rho_o^2-2\rho\rho_o\gamma)-\rho\rho_o\alpha(\lambda,\gamma) \left[  \frac{\left(\frac{\rho-\rho_o}{\rho+\rho_o}\right)^{2\alpha(\lambda,\gamma)}-2\pi\nu}{\left(\frac{\rho-\rho_o}{\rho+\rho_o}\right)^{2\alpha(\lambda,\gamma)}+2\pi\nu} \right]   \Bigg\}~,
\end{align}
while the Ricci scalar is given by
\begin{align}
     \mathcal{R}=&(\g^2-1)\rho^4\rho_o^2 \frac{(\rho-\rho_o)^{2(\g-2)}}{(\rho+\rho_o)^{2(\g+2)}}\cdot\nn \\[0.2cm]
    & \frac{(3|\lambda|-2)\left[  \left(\frac{\rho-\rho_o}{\rho+\rho_o}\right)^{2\alpha(\lambda,\gamma)}  + 4\pi^2\nu^2 \left(\frac{\rho+\rho_o}{\rho-\rho_o}\right)^{2\alpha(\lambda,\gamma)}   \right] -4\pi\nu (15|\lambda|+2)  }{ 2\pi\abs{ \left(\frac{\rho-\rho_o}{\rho+\rho_o}\right)^{\alpha(\lambda,\gamma)}  + 2\pi\nu \left(\frac{\rho+\rho_o}{\rho-\rho_o}\right)^{\alpha(\lambda,\gamma)}            }  }\,.
\end{align}
One can observe the following:

    In general, the factor $(\rho-\rho_o)^{\alpha(\lambda,\gamma)-2(\gamma-1)}$ compels $r(\rho)$ to vanish at the point $\rho=\rho_o$ since, $\alpha(\lambda,\gamma)-2(\gamma-1)>0$ for every $\g^2<1$. The Ricci scalar is singular at $\rho=\rho_o$ and since $g_{tt}$ does not vanish anywhere at $\rho>\rho_o$ we conclude that this point corresponds to a naked singularity at the center of the spherical symmetry $\rho=\rho_o$ (or $r=0$). Note that the function $r(\rho)$ is almost always increasing in the range $\rho>\rho_o$ therefore $\rho\rightarrow\infty$ corresponds to the asymptotic region $r\rightarrow\infty$.

    However, as we saw earlier, our solution contains Brans Class I as a special case, a family of solutions which is known to describe wormholes for some range of its parameters. With this in mind, it would be no surprise if our solution can describe wormhole spacetimes besides the naked singularities. A wormhole spacetime should contain two asymptotic regions and a throat that connects them. Hence, the areal radius should diverge at a second point except the one at $\rho\rightarrow\infty$ and additionally should have a minimum value which is greater than zero. The only possible case for which \eqref{arealradius,e>0} produces a second asymptotic region is if we demand the denominator to vanish i.e. $(\rho-\rho_o)^{2\alpha(\lambda,\gamma)}+2\pi\nu(\rho+\rho_o)^{2\alpha(\lambda,\gamma)}=0$. Solving this equation we can find the point $\rho$ at which the second asymptotic region occurs. It is given by
    \begin{align}
        \rho = \rho_o\frac{1+N}{1-N}\quad,\quad N=(-2\pi\nu)^{1/2\alpha}\label{e>0,point of 2nd asympt region}
    \end{align}
     and in order to be greater than $\rho_o$ the following relation must hold
     \begin{align}
         0<(-\nu)^{1/2\alpha}<(2\pi)^{-1/2\alpha}\,
     \end{align} which is satisfied irrespective of the values of $\lambda$ and $\gamma$ provided that $\nu$ lies in the range
     \begin{align}
         -\frac{1}{2\pi}<\nu<0\,.\label{e>0,n-bound for wormhole}
     \end{align}
    Again, to determine the redshift and shape functions we make use of \eqref{redshift-function}, \eqref{shape-function}. Their explicit forms are given in the appendix \ref{appendixA}. In the following diagrams (Fig. \ref{fig:e>0,wormhole-behaviour}, \ref{fig:e>0,energy-density}) we present the behaviour of the metric, areal radius, Ricci scalar and energy density $\varrho$ with respect to the isotropic radius.  In Fig. \ref{fig:e>0,BDvsCBD2} and \ref{fig:e>0,BDvsCBD} we compare the energy density of our model with the one obtained from Brans Class 1 solution for two different values of the parameter $\nu$. Both solutions violate the WEC near the wormhole throats.

\begin{figure}[H]
    \centering
    \includegraphics[scale=0.85]{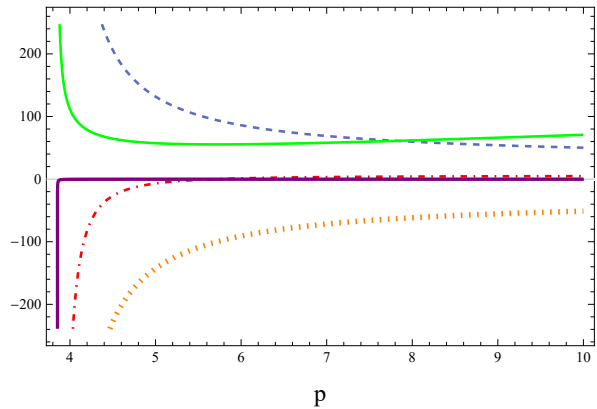}
    \caption{Plot of $g_{tt}$(orange-dashed line), $g_{\rho\rho}$(blue-dashed line), areal radius $r$(green solid line), $\frac{dr}{d\rho}$(red dashed-dotted line) and Ricci scalar(purple solid line) for parameter values $\rho_o=1,\gamma=0,\lambda=2,\nu=-0.019$ i.e. satisfying \eqref{e>0,n-bound for wormhole}. Areal radius diverges at two different points (one at $\rho\rightarrow\infty$ and one at the point given by \eqref{e>0,point of 2nd asympt region} ) corresponding to two different asymptotic regions. Between these two regions there is a point of minimum $r(\rho)$ where the throat is located.}
    \label{fig:e>0,wormhole-behaviour}
\end{figure}
\begin{figure}[H]
    \centering
    \includegraphics[scale=0.85]{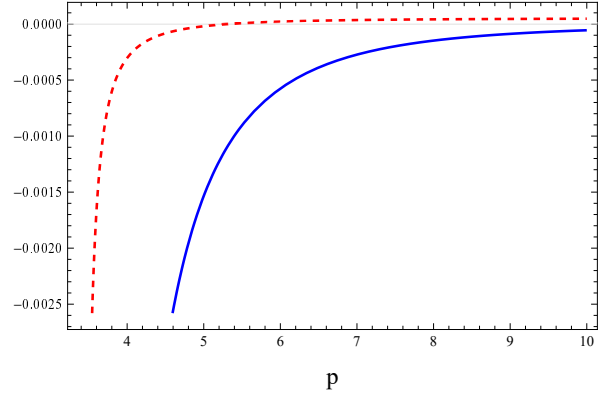}
    \caption{Plot of energy density$\varrho$(blue soldi curve) and  $\frac{dr}{d\rho}$(red dashed-dotted line) for $\rho_o=1,\gamma=0.5,\lambda=2,\nu=-0.019$ i.e. satisfying \eqref{e>0,n-bound for wormhole}. The energy density becomes negative as we approach the wormhole throat violating the WEC. }
    \label{fig:e>0,energy-density}
\end{figure}
\begin{figure}[H]
    \centering
    \includegraphics[scale=0.85]{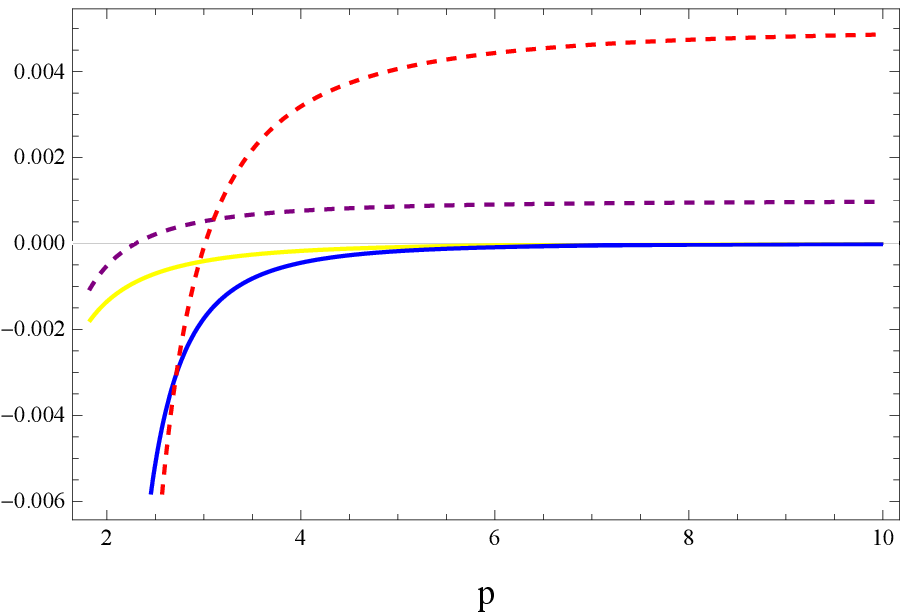}
    \caption{Plot of energy densities $\varrho$(blue solid curve), $\varrho_{BD}$(yellow solid curve) and  $\frac{dr}{d\rho}$(red dashed-dotted line), $\frac{dr_{BD}}{d\rho}$(purple dashed-dotted line) for $\rho_o=1,\gamma=0.5,\lambda=2,\nu\simeq -2.19\cdot10^{-3}$ i.e. satisfying \eqref{e>0,n-bound for wormhole}. The energy density becomes negative as we approach the wormhole throat for both theories violating the WEC. }
    \label{fig:e>0,BDvsCBD2}
\end{figure}
\begin{figure}[H]
    \centering
    \includegraphics[scale=0.85]{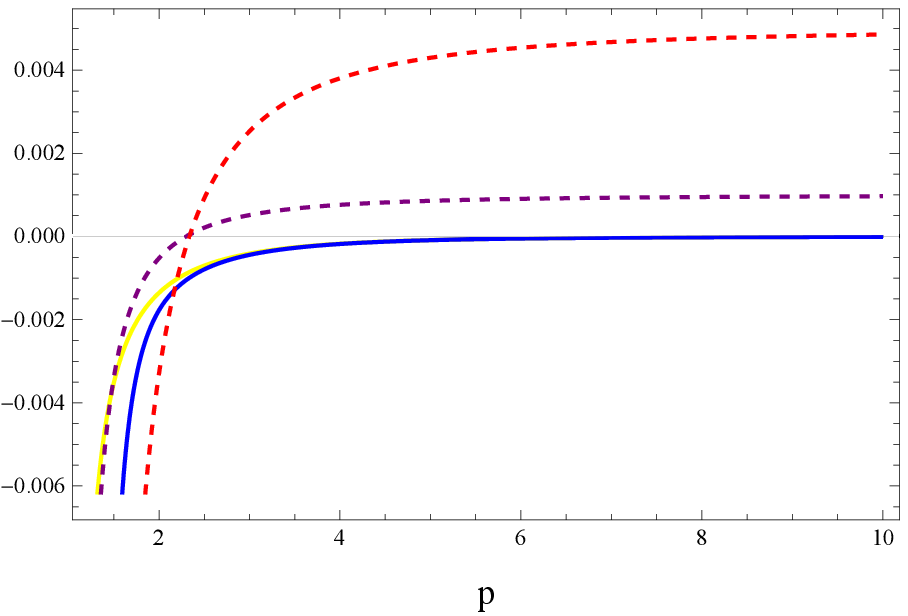}
    \caption{Plot of energy densities $\varrho$(blue solid curve), $\varrho_{BD}$(yellow solid curve) and  $\frac{dr}{d\rho}$(red dashed-dotted line), $\frac{dr_{BD}}{d\rho}$(purple dashed-dotted line) for $\rho_o=1,\gamma=0.5,\lambda=2,\nu\simeq -5.8\cdot10^{-4}$ i.e. satisfying \eqref{e>0,n-bound for wormhole}. The energy density becomes negative as we approach the wormhole throat for both theories violating the WEC. }
    \label{fig:e>0,BDvsCBD}
\end{figure}
    If $\gamma=1$ then
    \begin{align}
        r=\left(\frac{8\pi}{1+2\pi\nu}\right)^{1/2}\frac{(\rho+\rho_o)^2}{\rho}\,,\\[0.2cm]
        \frac{dr}{d\rho}=\left(\frac{8\pi}{1+2\pi\nu}\right)^{1/2} \left( 1-\frac{\rho_o^2}{\rho^2} \right)\,.
    \end{align}
    If $\gamma=-1$ we get
    \begin{align}
           r=\left(\frac{8\pi}{1+2\pi\nu}\right)^{1/2}\frac{(\rho-\rho_o)^2}{\rho} \,, \\[0.2cm]
        \frac{dr}{d\rho}=\left(\frac{8\pi}{1+2\pi\nu}\right)^{1/2} \left( 1-\frac{\rho_o^2}{\rho^2} \right)\, .
    \end{align}
    Hence, just like in case $\epsilon<0$, if $\gamma=\pm 1$ then the areal radius is an decreasing function for $0<\rho<\rho_o$, has an absolute minimum at $\rho=\rho_o$ and increases for $\rho>\rho_o$. If $\gamma=-1$ its minimum value is $r(\rho=\rho_o)=0$ whereas, if $\gamma=1$ then $r(\rho=\rho_o)=\left(\frac{8\pi}{1+2\pi\nu}\right)^{1/2}4\rho_o>0 $.

    In the limit $\nu\rightarrow0$ the areal radius yields
    \begin{align}
        r&=\sqrt{8\pi}\, \frac{(1+\rho_o/\rho)^{\alpha(\lambda,\gamma)+2(\gamma+1)}}{(1-\rho_o/\rho)^{\alpha(\lambda,\gamma)+2(\gamma-1)}}\,\rho\,,
    \end{align} and it is $\alpha
    (\lambda,\gamma)+2(\gamma-1)>0\longrightarrow\frac{2-|\lambda|}{2+|\lambda|}<\gamma<1$. If the $\gamma$ parameter lies within this range then  the point $\rho=\rho_o$ corresponds to $r\rightarrow\infty$. It's derivative is given by
    \begin{align}
        \frac{dr}{d\rho}=\frac{\sqrt{8\pi}}{\rho^2}\,\Bigg| \left(\frac{\rho+\rho_o}{\rho-\rho_o}\right)^{2\gamma+\alpha(\lambda,\gamma)} \Bigg|^{1/2}\,\left( \rho^2+\rho_o^2-2\rho\rho_o\gamma-\rho\rho_o\alpha(\lambda,\gamma) \right)\,
    \end{align}
    and the point of minimum $r$ corresponds to
    \begin{align*}
        \rho = \frac{\rho_o}{2}\left(2\gamma+\alpha(\lambda,\gamma)\pm\sqrt{-4+(2\gamma+\alpha(\lambda,\gamma))^2}\right)\,.
    \end{align*}
    By the use of \eqref{CDBtoBDparameters} we can express these roots in terms of the parameters of Brans Class I and the result is given by
    \begin{align}
        \rho = \frac{B(C+1)}{\lambda_I}\left( 1\pm \sqrt{1-\left(\frac{\lambda_I}{C+1}\right)^2} \right)^2~,
    \end{align} which is in agreement with the previously obtained result in \cite{Faraoni:2016ozb}.

\begin{figure}[H]
    \centering
    \includegraphics[scale=0.85]{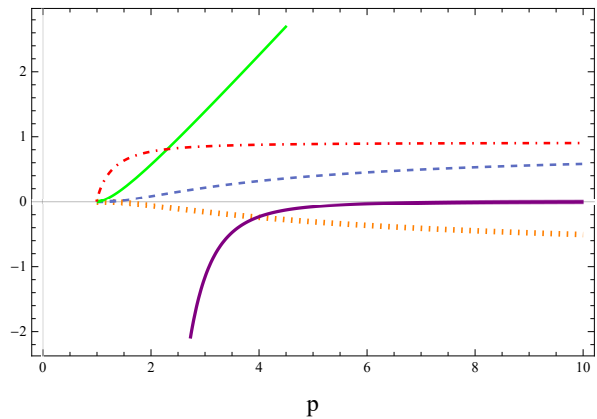}
\caption{Plot of  $g_{tt}$(orange-dashed line), $g_{\rho\rho}$(blue-dashed line), areal radius $r$(green solid line), $\frac{dr}{d\rho}$(red dashed-dotted line) and Ricci scalar(purple solid line) with respect to the isotropic radius for $\rho_o=1\,,\gamma=0.2\,,\lambda=2\,,\nu=-5$. At $\rho_o$ the areal radius as well as the metric components vanish whereas the Ricci scalar diverges. Unless $\gamma=\pm1$, the solution produces a singularity at $\rho=\rho_o$ which is not covered by a event horizon.}
\label{fig:allinplot}
\end{figure}

The behaviour of the metric, areal radius and Ricci scalar, for values of $\nu$ that do not satisfy \eqref{e>0,n-bound for wormhole}, is depicted in the graph Fig. (\ref{fig:allinplot}).

To summarise for the branch $\epsilon>0$, if $\gamma=\pm1$ the spherically symmetric solutions reduce to the Schwarzschild solution with a constant scalar field, while if $\gamma \neq 1$ then we have a naked singularity when the parameter $\nu$ does not satisfy the bound \eqref{e>0,n-bound for wormhole}, and a wormhole solution otherwise.

\section{Conclusions}
\label{Sec5}

The Brans classes I-IV of solutions of BD theory according to  Hawking’s theorem \cite{Hawking:1972qk} always describe
either wormholes or else horizonless geometries containing naked singularities, and they never describe black
holes. Spacetimes having naked singularities are unphysical in the sense that the initial value problem fails, leaving
the theory void of predictability and wormholes as the only remaining Brans solutions. The Brans solutions are vacuum solutions and
the BD scalar acts as the only form of effective matter. However,  the BD scalar can violate all of the energy conditions, therefore
it is no surprise that one can obtain wormholes as solutions of vacuum BD theory.

Wormholes are exotic objects which require the energy conditions to be violated. This can be understood because a  wormhole requires light rays which enter at one mouth and emerge from the other to have cross-sectional areas initially decreasing and then increasing. This conversion can be produced by gravitational repulsion which acts on the light rays passing near the throat and this can happen  provided  that in this region resides a negative energy density, as it is effectively guaranteed by the BD scalar field $\phi$ in the case of $\gamma>1$. Note that in  GR the properties required for the functions $\Phi(r)$ and $b(r)$ cause such constraints on the matter stress tensor as to make necessary the occupance of exotic matter, especially in the wormhole throat, where the absence of a horizon is required. In the BD theory the role of exotic matter is instead played, if $\gamma>1$ (or $\omega<-2$), by the scalar field $\phi$.

The aim of this work was to study spherically symmetric solutions and in particular wormhole solutions in a modified  BD theory. The BD theory is modified by introducing a dimensionful parameter $\nu$ \cite{Kofinas:2015nwa, Kofinas:2015sjz} in the kinetic coupling of the scalar field to gravity. This coupling gives an extra contribution to the matter content of the BD theory. Solving the coupled Einstein-Klein-Gordon equations
we found new  spherically symmetric  solutions which depend on the new parameter $\nu$. Demanding the absence of unphysical propagating modes, like ghosts, we find two branches of solutions.

The spherically symmetric solutions of the first branch for any choice of parameters and independently of the value of the new parameter $\nu$ reduce to the Schwarzschild black hole or to a naked singularity with a constant scalar field and  also give new wormhole solutions. The second branch of solutions, except the Schwarzschild solution for negative values of the parameter $\nu$, gives wormholes  whose throat size is inversely proportional to the coupling parameter $\nu$, and  a non-trivial scalar field. Also we checked the WEC for both branches and we found that they are violated in both cases.

 \section*{Acknowledgements}
We thank George Kofinas for many discussions and suggestions and Valerio Faraoni for his valuable remarks and comments.

\appendix
\section{Energy density and radial pressure}\label{appendixA}
In this appendix we present the explicit form for the radial pressure and the energy density with respect to the isotropic radius for both branches.\\[0.2cm]
$\epsilon<0$:
\begin{align*}
    \varrho =&-\frac{    \rho^4 \rho_{o}^2 \sqrt{\left| \nu \right| } \left(\frac{\rho+\rho_{o}}{\rho-\rho_{o}}\right)^{-2 \gamma }}{     2 \sqrt{2 \pi } (\rho-\rho_{o})^4 (\rho+\rho_{o})^4  } \sec \left(\alpha(\lambda,\gamma) \log \left(\frac{\rho-\rho_{o}}{\rho+\rho_{o}}\right)\right) \cdot \\
    &\Bigg\{-2  \gamma \alpha(\lambda,\gamma)  \sin \left(2 \alpha(\lambda,\gamma) \log \left(\frac{\rho-\rho_{o}}{\rho+\rho_{o}}\right)\right)+4 \left(\gamma ^2-1\right) \cos ^2\left(\alpha(\lambda,\gamma) \log \left(\frac{\rho-\rho_{o}}{\rho+\rho_{o}}\right)\right)+\\
    &\left(\gamma ^2-1\right) \left| \lambda \right|  \left(\cos \left(2 \alpha(\lambda,\gamma) \log \left(\frac{\rho-\rho_{o}}{\rho+\rho_{o}}\right)\right)-9\right)\Bigg\}    \;,
\end{align*}
\begin{align*}
    p_r =& \frac{   \rho^3 \rho_{o} \sqrt{\left| \nu \right| } \left(\frac{\rho+\rho_{o}}{\rho-\rho_{o}}\right)^{-2 \gamma } }{2 \sqrt{\pi } (\rho-\rho_{o})^4 (\rho+\rho_{o})^4}     \Bigg\{  -4 \sqrt{1-\gamma ^2} \sqrt{\left| \lambda \right| } \left(-\rho^2+\gamma  \rho \rho_{o}-\rho_{o}^2\right) \tan \left(\alpha(\lambda,\gamma) \log \left(\frac{\rho-\rho_{o}}{\rho+\rho_{o}}\right)\right)\\
    &-3 \sqrt{2} \left(\gamma ^2-1\right) \rho \rho_{o} \left| \lambda \right|  \tan ^2\left(\alpha(\lambda,\gamma) \log \left(\frac{\rho-\rho_{o}}{\rho+\rho_{o}}\right)\right)-2 \sqrt{2} \left(\gamma ^2-1\right) \rho \rho_{o}\Bigg\} \cdot\\
    &\cos \left(\alpha(\lambda,\gamma)\log \left(\frac{\rho-\rho_{o}}{\rho+\rho_{o}}\right)\right)~.
\end{align*}
$\epsilon>0$:
\begin{align*}
   \varrho=& \frac{\rho^4 \rho_o^2 \left(\frac{\rho+\rho_o}{\rho-\rho_o}\right)^{-2 \gamma } \left(\rho^2-\rho_o^2\right)^{-2 \alpha(\lambda,\gamma)}}{  4 \pi  (\rho-\rho_o)^4 (\rho+\rho_o)^4 \sqrt{\left(\rho^2-\rho_o^2\right)^{-2 \alpha(\lambda,\gamma)} \left(2 \pi  \nu  (\rho+\rho_o)^{2 \alpha(\lambda,\gamma)}+(\rho-\rho_o)^{2 \alpha(\lambda,\gamma)}\right)^2}   }\cdot \\
    &\Bigg\{    \left(\gamma ^2-1\right) \left| \lambda \right|  \Bigg(4 \pi ^2 \nu ^2 (\rho+\rho_o)^{4 \alpha(\lambda,\gamma)}-36 \pi  \nu  (\rho-\rho_o)^{2 \alpha(\lambda,\gamma)} (\rho+\rho_o)^{2 \alpha(\lambda,\gamma)}+\\
    &(\rho-\rho_o)^{4 \alpha(\lambda,\gamma)}  \Bigg)      -2  \gamma  \alpha(\lambda,\gamma) \left((\rho-\rho_o)^{4 \alpha(\lambda,\gamma)}-4 \pi ^2 \nu ^2 (\rho+\rho_o)^{4 \alpha(\lambda,\gamma)}\right)-\\
    &2 \left(\gamma ^2-1\right) \left(2 \pi  \nu  (\rho+\rho_o)^{2 \alpha(\lambda,\gamma)}+(\rho-\rho_o)^{2 \alpha(\lambda,\gamma)}\right)^2   \Bigg\}   \,,
\end{align*}
\begin{align*}
   p_r =& \Bigg\{ 4 \pi  (\rho-\rho_o)^4 (\rho+\rho_o)^4 \left(\rho^2-\rho_o^2\right)^{- \alpha(\lambda,\gamma)} \left(2 \pi  \nu  (\rho+\rho_o)^{2 \alpha(\lambda,\gamma)}+(\rho-\rho_o)^{2 \alpha(\lambda,\gamma)}\right) \Bigg\}^{-1}\cdot\\
   & \rho^3 \rho_o \left(\frac{\rho+\rho_o}{\rho-\rho_o}\right)^{-2 \gamma } \left(\rho^2-\rho_o^2\right)^{-2 \alpha(\lambda,\gamma)}\, \Bigg[       -2 \alpha(\lambda,\gamma) \left(-\rho^2+\gamma  \rho \rho_o-\rho_o^2\right)\cdot\\
   &\left((\rho-\rho_o)^{4 \alpha(\lambda,\gamma)}-4 \pi ^2 \nu ^2 (\rho+\rho_o)^{4 \alpha(\lambda,\gamma)}\right)+\\
   &2 \left(\gamma ^2-1\right) \rho \rho_o \left(2 \pi  \nu  (\rho+\rho_o)^{2 \sqrt{2} \sqrt{1-\gamma ^2} \sqrt{\left| \lambda \right|
   }}+(\rho-\rho_o)^{2 \alpha(\lambda,\gamma)}\right)^2+\\
   &3 \left(\gamma ^2-1\right) \rho \rho_o \left| \lambda \right|  \left((\rho-\rho_o)^{2 \alpha(\lambda,\gamma)}-2 \pi  \nu  (\rho+\rho_o)^{2 \alpha(\lambda,\gamma)}\right)^2    \Bigg]~.
\end{align*}
where $\alpha(\lambda,\gamma)=\sqrt{2|\lambda|(1-\gamma^2)}\,$.


\end{document}